# *QuantEYE:* The Quantum Optics Instrument for OWL

*D. Dravins[1], C. Barbieri[2], R. A. E. Fosbury[3], G. Naletto[4], R. Nilsson[5], T. Occhipinti[6], F. Tamburini[7], H. Uthas[8], L. Zampieri[9]*

[1] Lund Observatory, Box 43, SE-22100 Lund, Sweden; E-mail: dainis@astro.lu.se

[2] Department of Astronomy, University of Padova, Vicolo dell'Osservatorio 2, IT-35122 Padova, Italy; E-mail: cesare.barbieri@unipd.it

[3] Space Telescope-European Coordinating Facility & European Southern Observatory, Karl-Schwarzschild-Straße 2, DE-85748 Garching bei München, Germany; E-mail: rfosbury@eso.org

[4] Dept. of Information Engineering, University of Padova, Via Gradenigo, 6/B. IT-35131 Padova, Italy; E-mail: naletto@dei.unipd.it

[5] Lund Observatory, Box 43, SE-22100 Lund, Sweden; E-mail: ricky@astro.lu.se

[6] Dept. of Information Engineering, University of Padova, Via Gradenigo, 6/B, IT-35131 Padova, Italy; E-mail: tommaso.occhipinti@dei.unipd.it

[7] Department of Astronomy, University of Padova, Vicolo dell'Osservatorio 2, IT-35122 Padova, Italy; E-mail: tamburini@pd.astro.it

[8] Lund Observatory, Box 43, SE-22100 Lund, Sweden; E-mail: helena@astro.lu.se

[9] INAF – Astronomical Observatory of Padova, Vicolo dell'Osservatorio 5, IT-35122 Padova, Italy; E-mail: zampieri@pd.astro.it

**Abstract:** *QuantEYE* is designed to be the highest time-resolution instrument on ESO:s planned *Overwhelmingly Large Telescope*, devised to explore astrophysical variability on microsecond and nanosecond scales, down to the quantum-optical limit. Expected phenomena include instabilities of photon-gas bubbles in accretion flows, p-mode oscillations in neutron stars, and quantum-optical photon bunching in time. Precise timescales are both variable and unknown, and studies must be of photon-stream statistics, e.g., their power spectra or autocorrelations. Such functions increase with the *square* of the *intensity*, implying an enormously increased sensitivity at the largest telescopes. *QuantEYE* covers the optical, and its design involves an array of photon-counting avalanche-diode detectors, each viewing one segment of the OWL entrance pupil. *QuantEYE* will work already with a partially filled OWL main mirror, and also without [full] adaptive optics.

## 1. INTRODUCTION

The frontiers of astronomy have expanded through observational breakthroughs. In the recent past, one major thrust in expanding parameter envelopes was the addition of new wavelength regions. Now that almost all regions are accessible, the thrust is moving towards higher spatial and temporal resolution. The latter require great light-collecting powers to obtain a meaningful signal within a small fraction of a second. *QuantEYE* is designed to be the highest time-resolution instrument on OWL, the planned *Overwhelmingly Large Telescope* of ESO, the *European Southern Observatory*.

## 2. HIGH-SPEED ASTROPHYSICS AND QUANTUM OPTICS

Numerous discoveries were made with resolutions of milliseconds and slower: optical and X-ray pulsars; planetary-ring occultations; rotation of cometary nuclei; cataclysmic variables; pulsating white dwarfs; flickering high-luminosity stars; X-ray binaries; gamma-ray burst afterglows, etc. A limit for such optical studies has been that CCD-like detectors do not readily permit frame-rates faster than 1 – 10 ms, while photon-counting detectors either had low quantum efficiency or else photon-count rates limited to no more than some hundreds of kHz. Parallel to such instrumental issues lies the limitation of telescope light-collecting power: for reasonable sensitivity there must be a photon flux to match the time resolution: microseconds require megahertz count rates.

*QuantEYE* on *OWL* is designed for sub-nanosecond time resolutions with GHz photon count-rates to match. This will enable detailed studies of phenomena such as: millisecond pulsars; variability close to black holes; surface convection on white dwarfs; non-radial oscillation spectra in neutron stars; fine structure on neutron-star surfaces; photon-gas bubbles in accretion flows; and possible free-electron lasers in the magnetic fields around magnetars. Besides such applications in high-speed astrophysics, the aim is to reach timescales sufficiently short to reveal the quantum-optical statistics of photon arrival times.

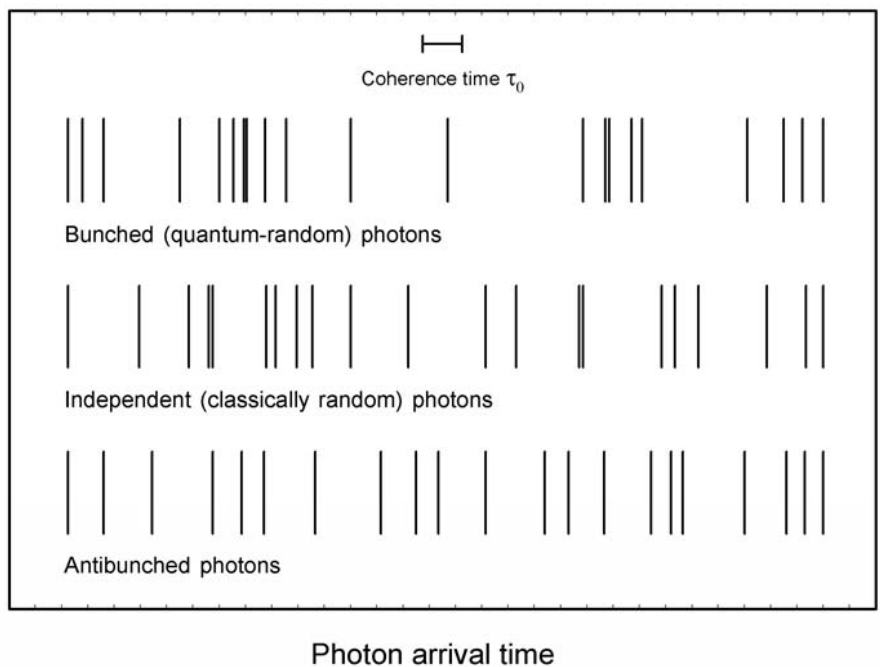

**Figure 1.** Statistics of photon arrival times in light beams with different entropies (different degrees or 'ordering'). Light may carry more information than that revealed by imaging and spectroscopy: Photons from given directions with given wavelengths give the same astronomical images and spectra, though the light may still differ in statistics of photon arrival times. These can be 'random', as in maximum-entropy black-body radiation (Bose-Einstein distribution with a certain 'bunching' in time), or may be quite different if the radiation deviates from thermodynamic equilibrium. (Adapted from Loudon, 2000)

### 2.1. Ultimate Information Content of Light

All existing astronomical instruments that record electromagnetic radiation of any wavelength are measuring either the directions of photon arrival (e.g., cameras), the energy of the arriving photons (e.g., spectrometers), or some combination of these properties. However, laboratory and theoretical studies in quantum optics have demonstrated that both individual photons and groups of photons may carry additional information, even for photons of some specific wavelength arriving from a given precise direction. Individual photons may carry various amounts of *orbital angular momentum* (in addition to their 'regular' angular momentum associated with circular polarization; e.g., Harwit 2003; Padgett et al. 2004), and the *statistics of photon arrival times* in an arriving photon stream may carry information on how the photon stream was created or was modified during its propagation (e.g.; Bachor 1998; Loudon 2000; Figure 1).

Current astronomical instrumentation exploits properties of the spatial coherence of light (for imaging) and of the temporal one (for spectroscopy). Beyond this first-order coherence, higher-order coherences of light may in principle convey information about the physics of light emission (e.g., stimulated emission as in a laser) or propagation (e.g., whether photons reach us directly from the source, or have undergone scattering on their way); Dravins 1994. In the laboratory, such properties of light have now been studied for some time, but have not yet been applied in astrophysics. Such higher-order coherence of light can be measured from the arrival-time statistics of individual photons, with the effects fully visible over timescales equal to the [first-order] temporal coherence time. For astronomically realistic passbands (1 nm, say), this is on the order of picoseconds, much shorter than realistic photometric resolutions. On the more realistic nanosecond scales, the effects are diluted but still measurable, as demonstrated already years ago by the intensity interferometer, the [so far] only astronomical instrument that studied the second-order coherence of light (e.g., Hanbury Brown 1974).

Using concepts related to this [spatial] intensity interferometer, analogous 'interferometry' in the time domain ('photon-correlation spectroscopy', also used in many laboratory experiments; e.g., Pike 1974) will reach a spectral resolution $\lambda/\Delta\lambda \approx 100,000,000$, as required to resolve known optical laser emission around the luminous object Eta Carinae. Theoretically expected emission linewidths there are on the order $\Delta\nu \approx 10$ MHz (Johansson & Letokhov 2004; 2005), which can be resolved by photon-correlation spectroscopy with delay times $\Delta t \approx 100$ ns. However, analogous to spatial information from intensity interferometry, photon correlation spectroscopy does not reconstruct the full *shape* of the source spectrum, but "only" gives linewidth information.

## 3. THE NEED FOR EXTREMELY LARGE TELESCOPES

The largest optical telescopes offer *enormously* increased sensitivity for studying astrophysical variability on timescales of milli-, micro-, and nanoseconds. Since the astrophysical phenomena are normally not periodic, and their exact timescales are both unknown and variable, studies must be of photon-stream statistics, e.g., power spectra or autocorrelations. Such functions increase with the *square* of the collected light intensity: doubling the telescope diameter increases the area fourfold, and the signal by a factor of 16! Higher-order correlations increase even steeper with telescope size.

The enormously increased sensitivity offered by ELT's in observing very rapid variability and photon statistics in astronomical sources, may well open up quantum optics as a fundamentally new information channel from the Universe. *QuantEYE* is being designed to meet that challenge, and to venture into previously unexplored parameter domains.

| Telescope diameter | Intensity $\langle I \rangle$ | Second-order correlation $\langle I^2 \rangle$ | Fourth-order photon statistics $\langle I^4 \rangle$ |
|---|---|---|---|
| 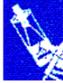 3.6 m | 1 | 1 | 1 |
| 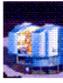 8.2 m | 5 | 27 | 720 |
| 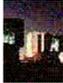 4 × 8.2 m | 21 | 430 | 185,000 |
| 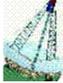 50 m | 193 | 37,000 | 1,385,000,000 |
| 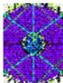 100 m | 770 | 595,000 | 355,000,000,000 |

**Table 1.** Light-curves become useless for resolutions below microseconds where typical time intervals between successive photons may even be longer than the time resolution. Instead studies have to be of the statistics of the arriving photon stream, such as its correlations or power spectra. All such statistical functions depend on [at least] the second power of the source intensity. The table compares the observed signal (I), its square and fourth powers, for telescopes of different size. The signal for classical quantities increases with the intensity I; the signal in power spectra as $I^2$; and that of four-photon correlations as $I^4$. This very steep dependence makes the largest telescopes enormously more sensitive for high-speed astrophysics and quantum optics.

## 4. *QuantEYE* CONCEPTUAL DESIGN

To study timescales down to nanoseconds, there is the corresponding need to count photons at sustained rates up to some GHz. The requirement of a high quantum efficiency leads to single-photon counting avalanche diodes (SPAD's) as the detectors of choice, although – at least at present – there appears not to exist any *single* detector that can handle such count rates, why we are led towards a concept of a detector *array*, over which the light from the source is distributed. A further technical limit is set by the – at least at present – small physical size (of order 100 µm) of the detector elements, which somewhat complicates the optical interface at large telescopes.

For the first conceptual design, a 'conservative' approach was taken, designing the system within existing detector technologies. Besides demonstrating the feasibility of concept, this means that a prototype or test instrument could be constructed along these lines, using commercially available components.

The optical design uses pupil-slicing, optically subdividing the OWL 100-meter entrance pupil into one hundred 10-meter segments. Light from these 100 pupil segments is then focused onto a fast (f/1) lenslet array with 100 lenses, feeding an array of 100 SPAD's through optical fibers. Each detector can sustain photon-count rates of up to some 10 MHz, so that their combined output may reach 1 GHz. Although, after photon detection, each detector has a deadtime of maybe 50 ns, the timing of each photon can be made with subnanosecond precision, as can the correlation between photon arrivals in different detectors. An exact differential timetag is assigned by a hydrogen maser clock (or future optical clock), and a GPS (or future *Galileo*) satellite receiver system provides an absolute time reference, enabling coordinated observations with other instruments on the ground or in space. A second detector head allows calibration and reference measurements.

Besides enabling GHz count rates, the segmented-pupil design has advantages in that (a) The detector redundancy enables to confirm possibly doubtful signals through their possible simultaneous occurrence in different channels; (b) Some types of events imply an illumination sweeping across the entrance pupil (e.g. occultations by Kuiper-belt asteroids), which now can be both spatially and temporally resolved; and (c) By suitable cross-correlations of the detected signal, a digital Hanbury Brown-Twiss intensity interferometer is realized between a large number of different sub-apertures across the full OWL pupil.

The *QuantEYE* wavelength range is set by that of its photon-counting SPAD's. The silicon-based ones cover the optical from 400 – 1000 nm, while for the near infrared (1.0 – 1.8 μm), SPAD's based on germanium and similar materials are being developed in industry. Such already exist, although their rather high dark-count rates do not yet make them suitable for our applications.

Raw data rates of 100 – 1000 Mb/s will be highly compressed in real time by on-line digital signal processors outputting only various statistical functions. Thanks to the pupil-slicing concept, *QuantEYE* will be able to work also with a partially filled OWL pupil, and (assuming the source is kept within the 1 arcsec aperture) will function well also without [full] adaptive optics.

Figure 2 outlines the optical solution for this conceptual design; for more details see Dravins et al. (2005):

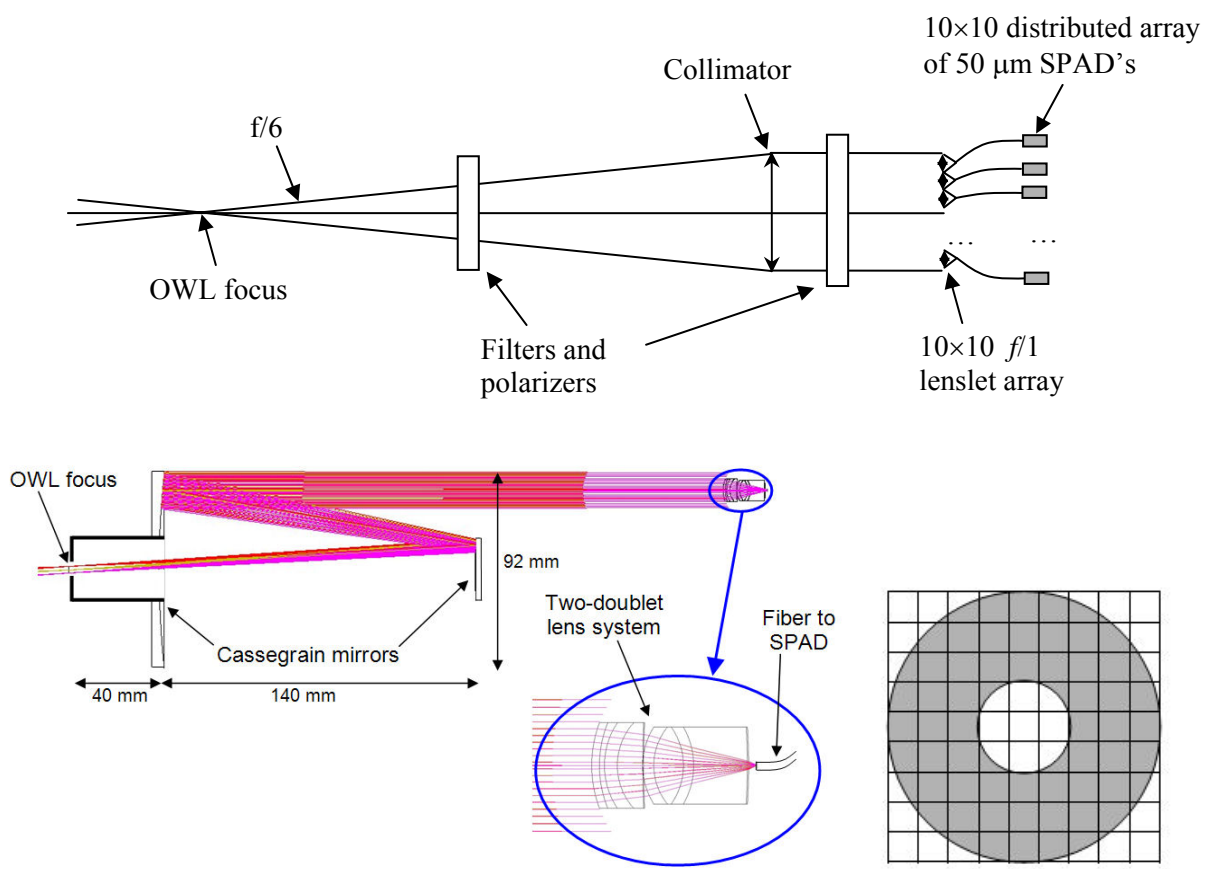

**Figure 2.** Sketch of the optical concept for *QuantEYE*: distributed detector array, and segmented aperture. The collimator-lens system magnifies 1/60 times (collimator focal length = 600 mm, lens focal length = 10 mm), giving a nominal spot size of 50 mm for a 1-arcsec source.

## 5. FUTURE DESIGN CHALLENGES

Removing the constraint to design the instrument with currenty available components, one may speculate what technology might eventually become available in the OWL construction timeframe.

The perhaps main limitation of the present design is that it only permits observations of one point source at a time. Of importance would be the availability of two-dimensional photon-detector arrays of high quantum efficiency, in which each pixel could sustain very high count rates (and ideally even have an energy/wavelength resolution). Such arrays are actively being developed, and if/when their performance reaches satisfactory levels, these should enable an *imaging* system with nanosecond resolution. For example, one could then observe a globular cluster containing an active X-ray source of unknown location, and then search for an optically rapidly variable object somewhere in the field.

The data rates could become quite significant. *If* an imaging system with one megapixel, say, becomes available, and if each pixel is photon-counting at 10 MHz, say, this generates $10^{12}$ time-tags per second, and petabytes of data already in a short time, stressing the need for efficient real-time data reduction, presumably outputting only the relevant statistical functions from each pixel, with a further on-line selection of the most 'interesting' pixels. In some sense, such a data 'filtering' is reminiscent of that being designed for high-energy accelerator experiments with analogous very large data flows.

## 6. INSTRUMENTATION PHYSICS

To eventually obtain high accuracies in astrophysical observations, a number of instrumental and observational issues may need to be understood. Here we exemplify some of them that merit further study:

* *Temporal structure of the sky background?* The night sky is not constant, and has contributions from, e.g., nanosecond-duration flashes of Cherenkov light induced by high-energy gamma rays (otherwise being studied by large light collectors), or faint but numerous meteors.

* *Atmospheric intensity scintillation*? Although, to a first approximation, the large OWL aperture averages out scintillation effects, contributions come due to the incompletely filled entrance pupil. The narrow interspaces between the mirror segments will extract the spatially smaller and temporally faster fluctuations (in addition being sensitive not only to scintillation amplitude, but also to the atmospheric wind direction, since the spaces between the hexagonal mirrors have only a few preferred directions; Dravins et al. 1998).

* *Temporal structure of stray light?* Some scattered light is always present, but normally it is not an issue whether it reaches the focal plane some nanoseconds sooner or later. However, if observing a sudden burst from a source, and if its light is scattered off some telescope structure, that scattered light may arrive to the detector with a systematic timelag relative to the main burst, possibly mimicking an afterpulse or a precursor (the speed of light is 30 cm per nanosecond).

* *Intensity fluctuations by adaptive optics?* The accurate photometry of [constant] objects across adaptive-optics corrected fields of view is being discussed in the literature. Any rapidly moving adaptive mirror will likely modulate the intensity of the source observed (at least changing some of the intensity fluctuations caused by the atmosphere). Since the adaptive-optics correction varies across the field, a calibration source elsewhere will probably not undergo the same modulation.

\* *Microphysics of photon detection*?  Contrary to common language usage, *photons* as such are never directly detected – rather one studies photo-*electrons* that result from the photon interaction inside the detector.  Photons, having integer quantum spin, are bosons and follow Bose-Einstein statistics which is very 'permissive' in allowing them to 'bunch' together in time.  Electrons, however, having half-integer spin, are fermions, follow Fermi-Dirac statistics, and obey the Pauli exclusion principle that prohibits any 'bunching' in the same quantum state.  One may then wonder how it is possible to study boson properties through a medium (electrons) that cannot, not even in principle, carry such properties?  For photocathode detectors, the explanation apparently is the very short time (femtoseconds?) required for a photoelectron to exit a photocathode and then be detected as an individual particle.  However, semiconductor detectors may be more complex, and have longer timescales for the relaxation of their inner energy levels.

\* *Quantum statistics changes inside the instrument?*  Photon statistics may be affected by many factors, including the re-direction of light, for example through a beamsplitter.  While in the classical case, a beamsplitter would split the amplitude of a light wave as 50-50%, say; in the quantum case the beamsplitter is 'cleaving' the photon gas with all its photon correlations into two parts whose statistics, i.e. the amount of photon bunching in time, is changed after passage of the beamsplitter (e.g., Bachor 1998).  On one hand, this illustrates how photon statistics is carrying information beyond the classical case, but it also demonstrates the need to understand the quantum behavior of the {telescope + instrument}-system: in the quantum world the observer may not be separable from what is being observed.

**Acknowledgements**

The work in Lund was supported by the Swedish Research Council, the Royal Physiographical Society in Lund, and the European Southern Observatory.  The work in Padova was supported by the University of Padova, INAF, and the European Southern Observatory.  At Lund Observatory we had valuable discussions with, in particular, Daniel Faria, Lennart Lindegren, and Hans-Günter Ludwig.  In the preliminary optical design, work by also F. Cucciarrè and V. Da Deppo (Dept. of Information Engineering, University of Padova) is acknowledged.